\documentclass[11pt,twoside]{article}
\usepackage{newpasp}

\index{Englmaier, P.}
\index{Rigopoulou, D.}
\index{Mengel, S.}

\begin{document}

\title{Gas Flow and Star Formation in the `Antennae' Galaxies}
\author{P. Englmaier, D. Rigopoulou, and S. Mengel}
\affil{MPI f\"ur extraterrestrische Physik, Garching, Germany}

\begin{abstract}

The prominent interacting galaxy pair NGC 4038/9 contains many active
star-forming regions and is continuously forming new star clusters.
We present a self-consistent n-body model for this system
which includes an SPH gas component. The model qualitatively explains
the apparent concentration of gas in the so-called overlap region {\em
between} the two nuclei as a bridge of gas connecting the two
galaxies. Projected on the sky, the bridge appears as a dense spot of
gas.  We discuss some implications for the evolution of Ultra-luminous
infrared galaxies.
\end{abstract}

The stellar dynamics of the Antennae galaxies merger has been studied
in detail by Toomre \& Toomre (1972), Barnes (1988), and others.  We
are interested in this merger to study star formation under extreme
conditions.  The gas transport towards the galaxy centers in the early
stages of the mergers may lead to circum-nuclear starbursts. Tidal
forces during the first passage trigger the formation of bars or
induce $m=2$ spiral arm modes in the approaching galaxies. Once a bar
is forming, the gas piles up at the inner Lindblad resonance which in
turn undergoes rapid star formation. The distribution of young stellar
clusters, is therefore linked to the dynamical history of the merger.

With bolometric luminosities and space densities comparable to or even
higher than those of quasars, Ultra-luminous Infrared Galaxies
(ULIRGs) are the most luminous objects (L$\geq$10$^{12}$L$_{\odot}$)
in the local Universe (Soifer et al. 1987). Deep ground based
(Rigopoulou et al. 1999) and HST--NICMOS (Scoville et al. 2000) images
reveal that ULIRGs are undergoing a major merger. Often the merging
includes two similar size disk galaxies.  However,
the merging sequence for ULIRGs seems to be somewhat different from
that of lower luminosity nearby mergers, implying that perhaps ULIRGs
have a different type of progenitor galaxies or their gas properties
are different.

Rigopoulou et al. (1999) have shown that in fact the gas properties of
ULIRGs are different than those of the less luminous
(L$\geq$10$^{11}$L$_{\odot}$) LIRG systems.  They found that although
in LIRGs the CO content decreases with decreasing separation, 
ULIRGs appear to still be gas-rich even at advanced stages of the
merger.  Activity in ULIRGs is known to be in part due to star bursts
and also due to AGN activity.  One may speculate, that AGN activity is
triggered by the merger event.  However, no correlation between AGN
activity and merger state has been found (Rigopoulou et al. 1999).  To
explain such results we have embarked into a project in investigating
whether the ULIRG progenitor galaxies are more stable against bar
formation than other mergers (Mihos \& Hernquist 1996) e.g. due to a
stronger bulge component.

While not being ULIRGs themselves, the Antennae galaxies provide an
excellent starting point for studying ULIRGs.  As a first step, we
model the Antennae system as closely as possible to explain gas
distribution and present star formation rate.  A nice goal is to
reproduce the observed distribution of young star clusters, which are
concentrated towards the so-called overlap region and also follow some
spiral arm structure. Our model presented here is similar to the model
by Barnes (1988), but in addition includes a gaseous component. The
n-body representation of both galaxies consists of four
components. The mass ratio for bulge:disk:halo:gas is 1:4:23:0.8. The
total number of particles is 100'000 (30'000 in gas).
The gas is treated as a single component SPH gas initially
distributed in a constant surface density disk. No attempt has been
made to improve the orbital parameters, as the match between the
observed and modeled distribution of stars is already quite good. Star
formation in the code is turned off.

The overlap region, where the two disks appear to overlap in
projection, corresponds in the model to a bridge between the galaxies
formed by $\sim20\%$ of the gas and $\sim30\%$ of the stars. The gas
fraction is likely to increase when an exponential gas disk is used.
In fact, both galaxies are distorted by tidal forces into bars aligned
with each other in such a way that they form a connected V-shaped
structure with the kink in the overlap region.  Observations indicate
a large amount of gas, dust and ongoing star formation
in this region (Whitmore et al. 1999; Mengel et al. 2000). From the time
sequence of our merger simulation, we conclude that the gas in the
overlap region originates mostly from inside the disk region where the
bars are formed by tidal forces. But some gas also comes from the
region which was compressed first in the collision between the two
disks, i.e. from larger disk radii.

Since the formation of the bars further amplifies the compression, it
may not be surprising that most young star clusters are found in that
area. When seen from the side, the gaseous bridge is offset by a few
kpc from the stellar bars. In the model this is the time when the
separation between gas and stars is largest.  To understand the
formation of clusters in detail, a more realistic model for the gas is
required. Most likely, clusters are formed out of giant molecular
clouds which collapse from overpressure generated by the starburst
(Jog \& Solomon 1992). Therefore the number of star clusters formed
will also depend on the properties of the in falling galaxies.

\end{document}